\documentclass[aps, prd, nofootinbib, reprint, preprintnumbers, superscriptaddress, showkeys]{revtex4-2}

\usepackage{amsmath}
\usepackage{amssymb}
\usepackage{tabularx}
\usepackage{graphicx}
\usepackage{float}
\usepackage[usenames,dvipsnames,svgnames]{xcolor}
\usepackage{hyperref}
\usepackage{nameref}
\usepackage{comment}
\usepackage[T1]{fontenc}

\hypersetup{dvips,dvipdfm,colorlinks=true,urlcolor=black,filecolor=magenta,linktoc=page,citecolor=red,linkcolor=blue,bookmarks=true}

\begin{document}

\title{A closer look at Viaggiu Holographic Dark Energy with Statefinder Hierarchy and Fractional Growth Diagnostics}

\author{Somnath Saha}
\email{sahasomnath847@gmail.com}
\affiliation{Department of Mathematics, Sree Chaitanya College, Habra 743268, West Bengal, India}


\begin{abstract}

\begin{center}
(Dated: \today)
\end{center}

We employ the statefinder hierarchy along with the fractional growth parameter to investigate the recently proposed noninteracting Viaggiu holographic dark energy (VHDE) model with the future event horizon as the infrared cut-off. By adjusting different values of the VHDE model parameter $\delta$, we illustrate the evolution trajectories of several important parameters, prticularly, the statefinder hierarchy $S_3^{(1)}$, $S_3^{(2)}$, $S_4^{(1)}$, $S_4^{(2)}$ and the fractional growth parameter $\epsilon$ against the redshift $z$ and the fractional energy density of matter, $\Omega_m$. Together with the growth rate of matter perturbations, we present the statefinder hierarchy that defines a composite null diagnostic (CND), which can differentiate dynamical dark energy from $\Lambda$CDM. The joint use of the statefinder hierarchy and the fractional growth parameter offers a powerful diagnostic tool for probing VHDE, effectively resolving the degeneracy between different parameter choices within the model. We find that employing a CND pair, rather than a single diagnostic, significantly enhances the diagnostic power for the VHDE model. Finally, we perform an $\omega_d-\omega_d'$ analysis and, in addition, assess the squared sound speed $v_{s}^{2}$ to determine the model's dynamical stability under small perturbations.
\keywords{}

\end{abstract}

\maketitle

\onecolumngrid
\section{\label{sec-1} Introduction}

\par Several independent cosmological observations \cite{SupernovaSearchTeam:1998fmf,Perlmutter1,WMAP:2003elm,SDSS:2003eyi,Planck:2013pxb,Planck:2015mvg} provide strong evidence that the Universe is presently experiencing an accelerated phase of expansion. This acceleration at late times is typically attributed to an unknown cosmic component, dubbed dark energy (DE), which is characterized by a large negative pressure and dominates the total energy density of the Universe, contributing nearly $70\%$. The second most significant contribution to the cosmic energy budget arises from dark matter (DM), estimated to be around $25\%$, the rest $5\%$ being ordinary matter, collectively known as baryons. Despite their crucial influence on the evolution of the Universe, the physical origin and fundamental properties of both DE and DM remain largely unexplained. The $\Lambda$CDM model, commonly considered the cornerstone of modern Cosmology, provides a compelling explanation for the formation of large-scale structures and the overall evolution of the Universe, showing strong agreement with a broad range of current observational evidence~\cite{Planck:2013pxb,Planck:2015fie}. In the $\Lambda$CDM framework, the cosmological constant ($\Lambda$), with an equation of state (EoS) $w_{\Lambda} = -1$, is responsible for the present accelerated expansion, acting as DE, while DM plays a central role in the formation of cosmic structures in the early Universe. Despite its successes, long-standing theoretical issues—such as the fine-tuning problem and the cosmic coincidence problem \cite{Weinberg:1988cp,Steinhardt:1999nw}—highlight unresolved questions regarding the nature of DE and why it becomes dominant only in the present cosmological epoch. In parallel, ongoing tensions in observational parameters, notably the $H_0$ tension \cite{Abdalla:2022yfr,DiValentino:2022uvj} and the $S_8$ tension \cite{Basilakos:2017rgc,DES:2017myr}, introduce further complications, challenging both theoretical models and the interpretation of cosmological data. For an up-to-date and thorough overview of the $\Lambda$CDM model, see \cite{2022NewAR..9501659P}. A broad range of theoretical models has been proposed to account for the observed acceleration, and detailed reviews of these approaches are available in the literature \cite{copeland2006dynamics,Bamba:2012cp,Capozziello:2011et,Nojiri:2010wj,Cai:2015emx,CANTATA:2021asi}. Nonetheless, the underlying mechanism responsible for the emergence and nature of cosmic acceleration continues to be one of the most challenging open problems in modern Cosmology.\\

An appealing framework for the quantitative modeling of DE is inspired by the holographic principle \cite{tHooft:1993dmi,Susskind:1994vu,fischler1998holography,cohen1999effective,Horava:2000tb,Bousso:2002ju}. In this approach, holographic dark energy (HDE) gives rise to a wide range of cosmological behaviors in both its simplest formulation and in various extended versions. These models are largely based on the identification of different cosmological horizons as the effective infrared cutoff, or characteristic length scale, of the Universe. Detailed discussions of horizon-based HDE models can be found in Refs.~\cite{li2004model,Horvat:2004vn,Pavon:2005yx,Wang:2005jx,huang2004holographic,Kim:2005at,Gong:2004fq,Nojiri:2005pu,Setare:2008hm,Granda:2008dk,Saridakis:2007cy,Setare:2006wh,Nojiri:2019kkp,Zhang:2009un,Lu:2009iv}. In this regard, it is pertinent to note that investigations carried out by the CANTATA Network \cite{CANTATA:2021asi} suggest that HDE frameworks do not suffer from many of the conceptual issues commonly encountered in modified gravity theories. A variety of HDE models can be constructed by generalizing the standard entropy--area relation underlying the holographic principle. These extensions include nonadditive entropy formulations proposed by Tsallis \cite{1988JSP....52..479T,Tsallis:2012js},  power-law corrections to the standard entropy \cite{Das:2007mj,Radicella:2010ss}, relativistic entropy introduced by Kaniadakis \cite{Kaniadakis:2002zz,Kaniadakis:2005zk}, and quantum-gravity--motivated modifications such as the Barrow entropy \cite{Barrow_2020} among others. Each of these generalized entropy prescriptions gives rise to a corresponding HDE model, namely Tsallis HDE \cite{Tavayef:2018xwx,Zadeh:2018poj,saridakis2018holographic,DAgostino:2019wko,Ghaffari:2020nnk}, power-law HDE \cite{Telali:2021jju}, Kaniadakis HDE \cite{Drepanou:2021jiv,Hernandez-Almada:2021aiw,Moradpour:2020dfm}, and Barrow HDE \cite{saridakis2020barrow} respectively. Notably, these extended entropy models typically introduce a parameter that quantifies deviations from the standard Bekenstein--Hawking entropy \cite{Bekenstein:1972tm,bekenstein1973black,hawking1975particle}. Furthermore, these and several other less common HDE variants have been studied in the literature in the context of different horizon scales as well as in scenarios involving interactions with DM \cite{Sheykhi:2011egx,Sadri:2019qxt,Bhattacharjee:2020rqk,Mamon:2020wnh,Mamon:2020spa,Adhikary:2024sax,Sheykhi:2025gkx}. For a comprehensive account of the HDE framework and its extensions, the reader may consult \cite{Wang:2016och} and the references therein. Finally, we note that a generalized form of horizon entropy has recently been proposed \cite{Nojiri:2024zdu,Nojiri:2023wzz,Odintsov:2024ipb,Luciano:2026ufu,Luciano:2025ovj,Leizerovich:2026pfy}, which connects the FLRW field equations of general gravity theories with the thermodynamics of the dynamical apparent horizon. The majority of the entropy forms outlined above can be derived as specific cases from this generalized entropy.\\

Using results related to black hole formation in a spatially flat, expanding Friedmann--Lemaître--Robertson--Walker (FLRW) universe, Viaggiu~\cite{Viaggiu:2014woa} proposed a modified form of the Bekenstein--Hawking entropy. This particular form of entropy was later investigated in a cosmological framework by evaluating it at the dynamical apparent horizon of the Universe~\cite{Viaggiu:2015cra}. Within this setting, a revised expression for the internal energy was obtained, which remains conserved throughout the cosmological evolution when measured at the apparent horizon. The presence of such a conserved quantity is in strong agreement with the holographic principle. Moreover, an analysis based on gravitational thermodynamics demonstrated that this extended entropy formulation forbids the emergence of a phantom-dominated era~\cite{Saha:2019qnx}. This conclusion differs significantly from results derived using the standard Bekenstein--Hawking entropy and is consistent with current observational constraints. Very recently, Saha et al. \cite{Saha:2026lnb} introduced Viaggiu holographic dark energy (VHDE) as an alternative approach to the quantitative description of DE. Based on the holographic principle and the use of Viaggiu entropy, this framework exhibits richer phenomenology and may provide a viable DE candidate \cite{Halder:2026wvg}.
 As the number of DE models continues to rise, there is a growing necessity for diagnostics that can distinguish between them. Several methods have emerged thus far, including the well-known statefinder \cite{Sahni:2002fz,alam2003exploring}, $Om$ \cite{sahni2008two}, and the growth rate of perturbations \cite{Acquaviva:2008qp,Acquaviva:2010vr,Wang:1998gt}. The statefinder offers a robust and sensitive geometrical diagnostic of DE, constructed using the second and third derivatives of the scale factor. Later, Arabsalmani and Sahni \cite{Arabsalmani:2011fz} further developed the statefinder by incorporating higher-order derivatives, referring to this new diagnostic as the `statefinder hierarchy'. Interested readers are referred to several papers that have employed the statefinder hierarchy and the growth rate of matter perturbations in various DE models \cite{Myrzakulov:2013owa,Li:2014mua,Zhang:2014sqa,Cui:2015ueu,Zhou:2016rtz,Mukherjee:2018oll,AlMamon:2021sgv,Bhardwaj:2021chg,Dubey:2025hsf}. 
 A comprehensive evaluation of the VHDE model necessitates further analysis using a range of cosmological diagnostics, including the statefinder hierarchy, the growth of large-scale structures, and the $\omega_d-\omega_d'$ plane. In this work, we employ these tools to gain deeper insight into the model’s cosmological behavior.
 Moreover, the dynamical viability of the model is examined by analyzing its response to small-scale perturbations through the squared sound speed, which serves as a crucial indicator of stability.\\

This paper is arranged as follows: In Sec. \ref{sec02}, we conduct a brief review of the non interacting VHDE model. In Sec. \ref{sec03} and Sec. \ref{sec04}, respectively, the statefinder hierarchy and the growth rate of perturbations are studied. Sec. \ref{sec05} presents $\omega_d-\omega_d'$ analysis and also the stability analysis using evolutionary trajectory of the squared sound speed is given. Finally, Sec. \ref{sec07} provides the conclusions and discussions.

\section{BRIEF REVIEW OF THE VIAGGIU HOLOGRAPHIC DARK ENERGY MODEL}
\label{sec02}

\par This section presents a concise overview of the theoretical formulation and cosmological implications of the VHDE scenario. Following the analysis of Viaggiu \cite{Viaggiu:2014woa}, the entropy of a black hole horizon is described by the modified expression 
\begin{equation}\label{eq-def-ve1}
S_{BH} =  \frac{\kappa_B}{4 L_p^2}A + \frac{3 \kappa_B}{2 c L_p^2} V H 
\end{equation}
Here, $A$ is the proper area of the black hole, and 
$L_p = {\left(\frac{\hbar G}{c^3}\right)}^{\frac{1}{2}}$ is the Planck length. $G$ is Newton's gravitational constant, $c$ is the speed of light in vacuum, $\hbar$ is the reduced Planck constant, and $\kappa_B$ is the Boltzmann constant. The quantity $V$ represents the effective geometric volume of the black hole, measured at its outer apparent horizon. Additionally, $H$ denotes the Hubble parameter. For a spatially flat FLRW universe, the entropy includes an extra work term (the second term in Eq. \ref{eq-def-ve1}), which serves as an indication of the dynamical degrees of freedom associated with the expanding Universe (H $\neq 0$). It is worth noting that the standard Bekenstein-Hawking entropy \cite{Bekenstein:1972tm,bekenstein1973black,hawking1975particle} is retrieved by setting $H = 0$ in the above formula. Since the area and volume enclosed by a horizon of radius $L$ are given by $4 \pi L^2$ and $\frac{4}{3} \pi L^3$, respectively, the Viaggiu entropy in Eq.~(\ref{eq-def-ve1}) can be expressed in terms of the length scale $L$ as:

\begin{equation}\label{eq-def-ve2}
S_{BH} =  \pi L^{2} + 2\pi H L^{3}
\end{equation}
From this point onward, we work in gravitational units where $G$, $c$, $\kappa_B$, and $\hbar$ are all set to 1.\\

The VHDE model, grounded in the modified entropy–area relation (\ref{eq-def-ve2}) and the HDE concept ($
M_p^{-2} L^4 \rho_{d} \leq S_{BH}, \quad \text{with} \quad M_p = {(8 \pi)}^{-\frac{1}{2}}$)~\cite{cohen1999effective,li2004model}, was proposed in \cite{Saha:2026lnb} by specifying the following form for the energy density:
\begin{eqnarray}\label{eq-rhodegenL}
\rho_{d} &=& \frac{\delta^{2}}{8\pi}  S_{BH} L^{-4} \nonumber \\
&=& \frac{\delta^2}{8}  (1 + 2 H L)L^{-2}.
\end{eqnarray}
Here, $\delta^2$ is a positive constant and is not required to be dimensionless.\\

Furthermore, we focus on a universe that is spatially flat, homogeneous, and isotropic, which can be described by the FLRW metric. The geometry of this spacetime is expressed by the line element:
\begin{equation}
ds^{2} = - dt^{2} + a^{2}(t) \left[ dr^{2} + r^{2} \lbrace d\theta^{2} +  \sin^{2}\theta \, d\phi^{2}\rbrace \right],
\end{equation}
where $t$ denotes the cosmic time, and $a(t)$ is the scale factor that characterizes the expansion of the Universe. 
We assume that the total energy density is contributed by DM and VHDE. Under these conditions, the dynamics of the Hubble parameter and the scale factor are described by the Friedmann and the acceleration equations:
\begin{equation}\label{eq-fe1}
3H^{2} = 8\pi  \rho_{T},
\end{equation}
\begin{equation}\label{eq-fe2}
2\dot{H} + 3H^{2} = -8\pi  p_{T},
\end{equation}
Here, the Hubble parameter is given by $H = \frac{\dot{a}}{a}$, where the dot denotes derivative with respect to cosmic time. The total energy density $\rho_{T}$ and total pressure $p_{T}$ are obtained by summing the individual contributions of DM and DE, namely $\rho_{T}=\rho_{m}+\rho_{d}$ and $p_{T}=p_{m}+p_{d}$.\\

In the absence of any interaction between the dark sectors, each component evolves independently and obeys its own conservation equation. Under this assumption, the continuity equation for the $i$-th cosmic fluid takes the form
\begin{equation}\label{eq-rhodot}
\dot{\rho}_{i} + 3H(1+w_{i})\rho_{i} = 0,
\end{equation}
where $\rho_i \in \{\rho_d, \rho_m\}$. Here, $\rho_i$ represents the energy density of the corresponding component, and $w_i =\frac{p_i}{\rho_i}$ denotes its EoS. For DM, the pressure is negligible, $p_{m} = 0$, which implies $w_{d} = 0$. Consequently, the energy density of DM evolves with the scale factor as
\begin{equation}
\rho_{m} = \rho_{m0} a^{-3},
\end{equation}
where $\rho_{m0}$ represents the present-day DM density and the current scale factor is normalized to $a_{0} = 1$. \\

The fractional energy densities of DM and VHDE are expressed in terms of $\Omega_{m}$ and $\Omega_{d}$, respectively, as
\begin{equation}\label{eq-Omegadmdef}
 \Omega_{m} \equiv \frac{\rho_{m}}{\rho_c} = \frac{H^{2}_{0}\Omega_{m0}a^{-3}}{H^2}, 
\end{equation}

\begin{equation}\label{eq-Omegadedef}
 \Omega_{d} \equiv \frac{\rho_{d}}{\rho_c}= \frac{\delta^2 \pi (1 + 2 H L)L^{-2}}{3H^2}.
\end{equation}

where the critical energy density is defined as $\rho_c = \frac{3H^2}{8\pi }$. With these definitions, Eq.~(\ref{eq-fe1}) can be rewritten as
\begin{equation}\label{eq-Omegaconstr}
\Omega_{m}+\Omega_{d}=1.
\end{equation}

Substituting Eq. (\ref{eq-Omegadmdef}) into Eq. (\ref{eq-Omegaconstr}) subsequently gives
\begin{equation}\label{eq-H1a}
\frac{H_{0}}{H(a)}
= \frac{a^{\frac{3}{2}}\sqrt{\left(1-\Omega_{d}(a)\right)}}{\sqrt{\Omega_{m0}}}\, .
\end{equation}
where \(H_0\) and \(\Omega_{m0}=(1-\Omega_{d0})\) are the current values of the Hubble parameter and the DM density parameter, respectively.\\

It is a widely recognized fact that the future event horizon exists solely within an accelerating universe and is defined as the integral \cite{faraoni2015cosmological}
\begin{equation} \label{re-def}
R_E=a(t)\int_{t}^{\infty} \frac{dt}{a(t)}=a\int_{a}^{\infty} \frac{da}{Ha^2}=a\int_{x}^{\infty} \frac{dx}{Ha}.
\end{equation}
In physical terms, the future event horizon represents the proper distance to the most distant event that a comoving observer will ever perceive \cite{faraoni2015cosmological}.\\

Subsequently, in accordance with \cite{Saha:2026lnb}, if we substitute $L= R_E$ into Eq. (\ref{eq-rhodegenL}), the energy density of VHDE is transformed into
\begin{equation}\label{rhod-re}
\rho_d=\frac{\delta^2}{8R_{E}^2}(1+2HR_E).
\end{equation}
Moreover, when no interaction between DE and DM is present, the EoS for DE is formulated as \cite{Wang:2005jx}
\begin{equation} \label{wd-new}
w_d=-\frac{\Omega_{d}^{'}}{3\Omega_d(1-\Omega_d)},
\end{equation}
where $\Omega_d$ is the fraction of the overall energy density linked to DE and the prime denotes derivative with respect to $x = \ln a$.\\

To comprehend $\omega_d$, it is vital to identify $\Omega_{d}^{'}$. Therefore, we start by calculating the derivative of the VHDE density with respect to cosmic time \cite{Saha:2026lnb}:
\begin{equation}
\dot{\rho}_d=\frac{\delta^2}{4R_{E}^{3}}\left[1+R_{E}^{2}(\dot{H}-H^2)\right].
\end{equation}
Thereafter, we calculate the temporal derivative of the fractional DE density \cite{Saha:2026lnb}:
\begin{equation}
\dot{\Omega}_{d} =\frac{2\pi \delta^2}{3H^2R_{E}^{2}}\left[-R_E(\dot{H}+H^2)+\frac{1}{R_E}-\frac{\dot{H}}{H}\right].
\end{equation}
It is more suitable to evaluate the above derivative in terms of $\mbox{ln}~a$, which provides  \begin{equation}\label{omd-dash}
\Omega_{d}^{'}=\frac{2\pi \delta^2}{3H^2R_{E}^{2}}\left[-R_E\frac{\dot{H}}{H}-HR_E+\frac{1}{HR_E}-\frac{\dot{H}}{H^2}\right].
\end{equation}
Introducing $\Omega_{d}^{'}$ in Eq. (\ref{rhod-re}) gives rise to the quadratic equation \cite{Saha:2026lnb}
\begin{equation} \label{quad-e}
(HR_{E})^2-\left(\frac{2 \pi \delta^2}{3\Omega_{d}}\right)HR_E-\frac{\pi \delta^2}{3\Omega_{d}}=0.
\end{equation}
In accordance with the argument presented in \cite{Saha:2026lnb}, one obtains
\begin{equation}\label{hre}
HR_E=\frac{\pi \delta^2}{3\Omega_d}\left(1+\sqrt{1+\frac{3\Omega_d}{\pi \delta^2}}\right).
\end{equation}
Using Eqns. (\ref{eq-fe1}) and (\ref{rhod-re}) we find \cite{Saha:2026lnb}
\begin{eqnarray} \label{omdd}
\frac{\Omega_{d}^{'}}{\Omega_{d}(1-\Omega_{d})}=\frac{\frac{6\Omega_d}{\pi \delta^2}+\left(1+\sqrt{1+\frac{3\Omega_d}{\pi \delta^2}}\right)}{(2-\Omega_d)\left(1+\sqrt{1+\frac{3\Omega_d}{\pi \delta^2}}\right)-\frac{\frac{3\Omega_d}{\pi \delta^2}(1-\Omega_d)}{\sqrt{1+\frac{3\Omega_d}{\pi \delta^2}}}}.
\end{eqnarray}
In addition, when Eq. (\ref{omdd}) is multiplied by $-\frac{1}{3}$, the EoS for VHDE is obtained as \cite{Saha:2026lnb}

\begin{eqnarray} \label{wd-f}
\begin{split}  
w_d & = -\frac{\Omega_{d}^{'}}{3\Omega_{d}(1-\Omega_{d})}\\
& = -\frac{1}{3}\left[\frac{\frac{6\Omega_d}{\pi \delta^2}+\left(1+\sqrt{1+\frac{3\Omega_d}{\pi \delta^2}}\right)}{(2-\Omega_d)\left(1+\sqrt{1+\frac{3\Omega_d}{\pi \delta^2}}\right)-\frac{\frac{3\Omega_d}{\pi \delta^2}(1-\Omega_d)}{\sqrt{1+\frac{3\Omega_d}{\pi \delta^2}}}}\right].
\end{split}
\end{eqnarray}
Finally, the deceleration parameter $q$ is obtained as \cite{Saha:2026lnb}
\begin{eqnarray}
q &=& \frac{1}{2}(1+3w_d \Omega_d) \nonumber \\
&=& \frac{1}{2}-\frac{\Omega_d \sqrt{1+\frac{3\Omega_d}{\pi \delta^2}} \left[\left(1+\frac{3\Omega_d}{\pi \delta^2}\right)^{\frac{3}{2}}+\left(1+\frac{9\Omega_d}{2\pi \delta^2}\right)\right]}{\left(1+\sqrt{1+\frac{3\Omega_d}{\pi \delta^2}}\right)\left[\frac{3\Omega_d}{\pi \delta^2}+(2-\Omega_d)\left(1+\sqrt{1+\frac{3\Omega_d}{\pi \delta^2}}\right)\right]}
\end{eqnarray}
\section{STATEFINDER DIAGNOSIS FOR THE VHDE MODEL}\label{sec03}
In this section, utilizing the statefinder hierarchy diagnostic tool, we shall examine the VHDE model in order to differentiate it from the standard $\Lambda$CDM model. As we are particularly interested in the late-time evolution of the Universe, we shall consider the Taylor expansion of the scale factor about $t_0$, the present epoch \cite{Arabsalmani:2011fz}:
\begin{eqnarray}
    \frac{a(t)}{a(t_0)} &=& \frac{1}{1+z} \nonumber\\
    &=& 1+ \sum_{n=1}^{\infty} [H_0(t-t_0)]^n\frac{A_n(t_0)}{n!},
\end{eqnarray}
where
\begin{equation}
    A_n=\frac{1}{H^na(t)}\frac{d^na(t)}{d(t)^n}, ~~n\in N.
\end{equation}
It is essential to highlight that the term $A_2 = -q$ denotes the deceleration parameter, while $A_3$ signifies the statefinder $``r" $\cite{Chiba:1998tc,Sahni:2002fz,Alam:2003sc} or the jerk $ ``j" $\cite{Visser:2003vq} parameter. Furthermore, $A_4$ and $A_5$ correspond to the snap $`s'$ and the lerk $``l"$, respectively (refer to, for example \cite{Visser:2003vq,Capozziello:2008qc,Dunajski:2008tg,Dabrowski:2005fg}, and the references therein).
In the case of the $\Lambda$CDM model $(\omega_\Lambda =-1)$, we have
\begin{eqnarray}
    A_2 &=& 1-\frac{3}{2}\Omega_m,\nonumber\\
    A_3 &=& 1, \\
    A_4 &=& 1-\frac{3^2}{2}\Omega_m ~~\mbox{etc.} \nonumber
\end{eqnarray}
where, $\Omega_m=\frac{2}{3}(1+q)$.
Thus, for $\Lambda$CDM, it follows that all the $A_n $ parameters can be represented as simple functions of the deceleration parameter $q$ or the matter's fractional density parameter $\Omega_m$.\\

We are now in a position to define the \textit{ statefinder hierarchy} $S_n$ as \cite{Arabsalmani:2011fz}
\begin{eqnarray}
    S_2 &=& \frac{3}{2}\Omega_m+A_2,\nonumber \\
    S_3 &=& A_3,\\ 
    S_4 &=& \frac{3^2}{2}\Omega_m + A_4 ~~\mbox{etc.} \nonumber
\end{eqnarray}
A notable feature of this diagnostic is that all of the $S_n$ parameters remain equal to one throughout the entire cosmic expansion in the $\Lambda$CDM model, i.e., $S_n|_{\Lambda CDM}=1$ and it establishes a set of null diagnostics characterizing the $\Lambda$CDM model when $n>2$. Consequently, utilizing $\Omega_m = \frac{2}{3}(1+q)$ for $\Lambda$CDM, the statefinder hierarchy $S_3^{(1)}$ and $S_4^{(1)}$ can be expressed as \cite{Arabsalmani:2011fz}:
\begin{eqnarray}
    S_3^{(1)} &=& A_3,\nonumber\\ 
    S_4^{(1)} &=& 3(1+q) + A_4.
\end{eqnarray}
In a similar manner, one can construct higher order statefinders. Furthermore, another group of statefinders can be formulated based on the original set as indicated in \cite{Arabsalmani:2011fz}:
\begin{equation}\label{S3} 
    S_3^{(2)} =\frac{S_3^{(1)}-1}{3(q-\frac{1}{2})}
\end{equation}
In $\Lambda$CDM cosmology, the initial statefinder $S_3^{(1)}$ is equal to $1$, whereas the subsequent statefinder $S_3^{(2)}$ is equal to 0. Consequently,
$\left\{S_3^{(1)}, S_3^{(2)}\right\} = \{1, 0\}$ provides a model-independent method for differentiating the $\Lambda$CDM model from other DE models \cite{Sahni:2002fz,Alam:2003sc}. Based on Eq. (\ref{S3}), it is straightforward to define the second element of the statefinder hierarchy as 
\cite{Chiba:1998tc,Arabsalmani:2011fz}
\begin{equation}\label{Sn} 
    S_n^{(2)} =\frac{S_n^{(1)}-1}{\alpha(q-\frac{1}{2})}
\end{equation}
where $\alpha$ represents an arbitrary constant. Correspondingly, the $S_4^{(2)}$ can be described as \cite{Arabsalmani:2011fz}
 \begin{equation}\label{S4} 
    S_4^{(2)} =\frac{S_4^{(1)}-1}{9(q-\frac{1}{2})}
\end{equation}
 Furthermore, the superscript $``(1)"$ serves to distinguish not only between $S_n$ and $S_n^{(1)}$, but also between $S_n^{(1)}$ and its derivative $S_n^{(2)}$. In $\Lambda$CDM, $S_n^{(2)}$ equals 0, and consequently, a set of null diagnostics for DE can be established as $\left\{S_n^{(1)}, S_n^{(2)}\right\} = \{1, 0\}$, which presents a model-independent approach to differentiate evolving DE models
from the cosmological constant. For $\alpha =3$, $\left\{S_n^{(1)}, S_n^{(2)}\right\} = \{r, s\}$ \cite{Alam:2003sc,Sahni:2002fz}. In the context of any DE model, we have \cite{Alam:2003sc}
\begin{eqnarray}  
    S_3^{(1)}&=&1+ \frac{9}{2}\omega_d(\omega_d+1)\Omega_d-\frac{3}{2}\omega_d'\Omega_d
   \end{eqnarray}
   \begin{equation}    
    S_4^{(1)}= 1-\frac{9}{4} \omega_{d} \Omega_{d}^{2} \left[3 \omega_{d}\left(\omega_{d}+1\right)-\omega_{d}^{\prime}\right]\\- \frac{3}{4} \Omega_{d} \left[\omega_{d}\left(21+39 \omega_{d}+18 \omega_{d}^{2}\right)\right]
- \frac{3}{4} \Omega_{d} \left[-\left(13+18 \omega_{d}\right) \omega_{d}^{\prime}+2 \omega_{d}^{\prime \prime}\right].
\end{equation} 
where the prime symbol indicates the derivative with respect to $x = \ln a$.\\
\begin{figure*}[ht]
     \centering
     \begin{minipage}{0.425\textwidth}
         \centering
         \includegraphics[width=\textwidth]{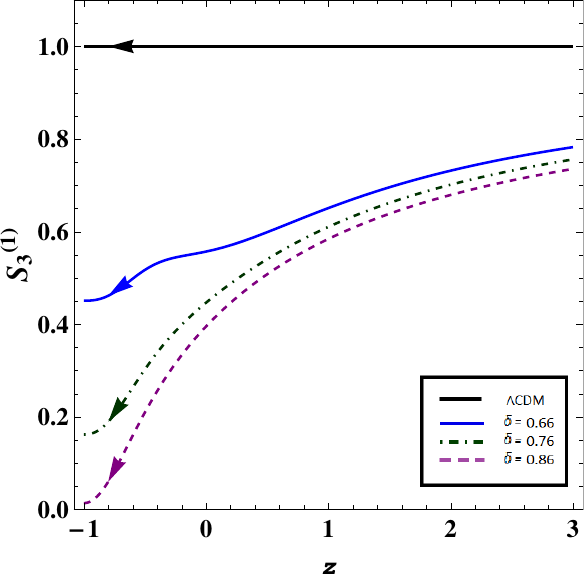}
     \end{minipage}
     \hspace{1cm}
     \begin{minipage}{0.425\textwidth}
         \centering
        \includegraphics[width=\textwidth]{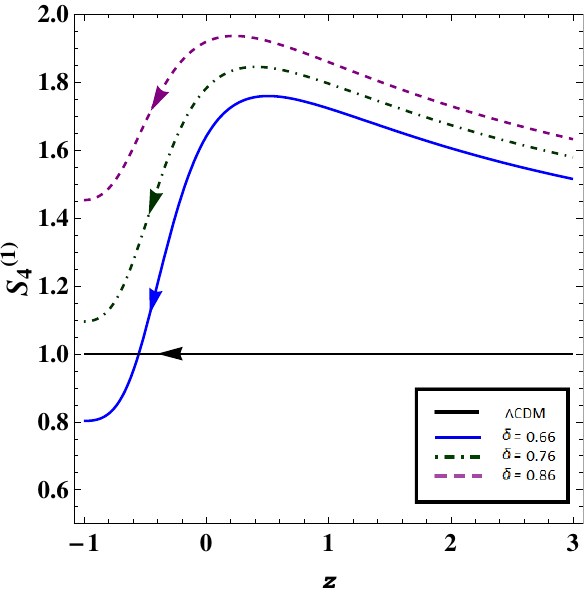}
     \end{minipage}
        \caption{The left and the right panels correspond to the evolutionary paths of $S_3^{(1)}(z)$ and $S_4^{(1)}(z)$, respectively, as a function of the redshift $z$, for different values of $\delta$. The arrows indicate the direction of cosmic evolution. The initial condition is considered as $\Omega_d(x=-\mbox{ln}(1+z)=0) \equiv \Omega_{d0} \approx 0.7$.}
        \label{fig 1}
\end{figure*}

FIG. \ref{fig 1} (left panel), shows the evolutionary paths of $S_3^{(1)}(z)$ in the VHDE for different values of
$\delta$. The separation of curvilinear shapes of the VHDE model  is more distinct in the region $-1 \leq z \leq 0$. It can be seen that all the curves $S_3^{(1)}(z)$ exhibit a similar behavior, they initially lie below the $\Lambda$CDM line $S_3^{(1)}=1$ and subsequently decrease monotonically from the high redshift region to low redshift region for different values of $\delta$.
All the curves 
follow the close degeneration 
in high redshift region.
The evolutionary paths of $S_4^{(1)}(z)$ for the VHDE model taking various values of $\delta$ are displayed in FIG. \ref{fig 1}. (right panel). We observe that for all values of $\delta$ the trajectories of $S_4^{(1)}(z)$ have close degeneration in the long past. They start evolving above $S_4^{(1)}=1$ line and grow monotonically from the high-redshift region. In the region $-1 \leq z \leq 1$  all curves form convex vertices and are well differentiated from each other at low red-shift region. These results show that different values of $\delta$ have quantitative impacts on $S_3^{(1)}(z)$ and $S_4^{(1)}(z)$.\\

\begin{figure*}[ht]
     \centering
     \begin{minipage}{0.425\textwidth}
         \centering
         \includegraphics[width=0.97\textwidth]{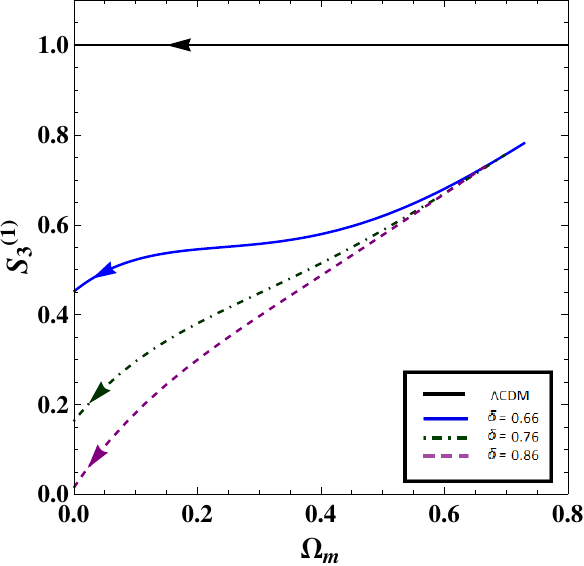}
     \end{minipage}
     \hspace{1cm}
     \begin{minipage}{0.425\textwidth}
         \centering
        \includegraphics[width=\textwidth]{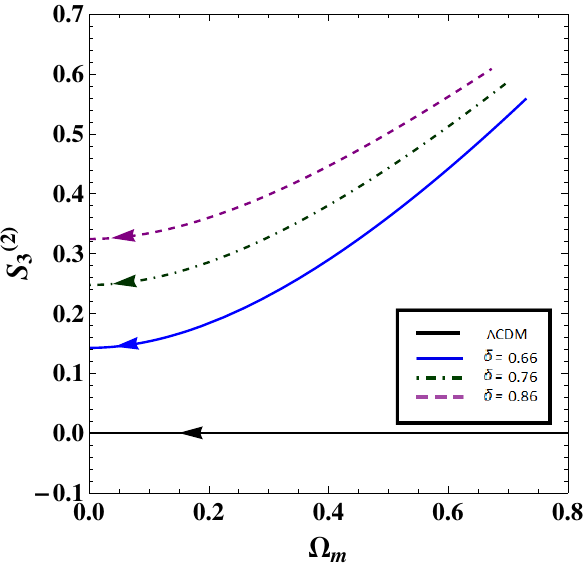}
     \end{minipage}
        \caption{The left and the right panels correspond to the evolutionary paths of $S_3^{(1)}$ and $S_3^{(2)}$, respectively, as a function of $\Omega_m$, for different values of $\delta$. The arrows indicate the direction of cosmic evolution. The initial condition is considered as $\Omega_d(x=-\mbox{ln}(1+z)=0) \equiv \Omega_{d0} \approx 0.7$.}
        \label{fig 2}
\end{figure*}

The evolutionary paths of $S_3^{(1)}$ and $S_3^{(2)}$ as a function of $\Omega_m$ are depicted in FIG. \ref{fig 2}.  Higher $\Omega_m$ values indicate the early universe ($z\gg 1$), while low $\Omega_m$ values, close to 0, are indicative of the future universe ($z$ approaching $-1$). We observe that for different values of $\delta$, the degeneracy of $S_3^{(1)}-\Omega_m$ curves (left panel) is perfectly broken in the present ($\Omega_m\approx0.3$) and future universe. The variation among the curves of $S_3^{(2)}-\Omega_m$ for different values of $\delta$ (right panel) are  minimal and all the trajectories decrease monotonically throughout the evolution.  It is worth noting that, for smaller values of $\delta$, the curves approach the $\Lambda CDM$ line more closely.\\

\begin{figure*}[ht]
     \centering
     \begin{minipage}{0.425\textwidth}
         \centering
         \includegraphics[width=0.97\textwidth]{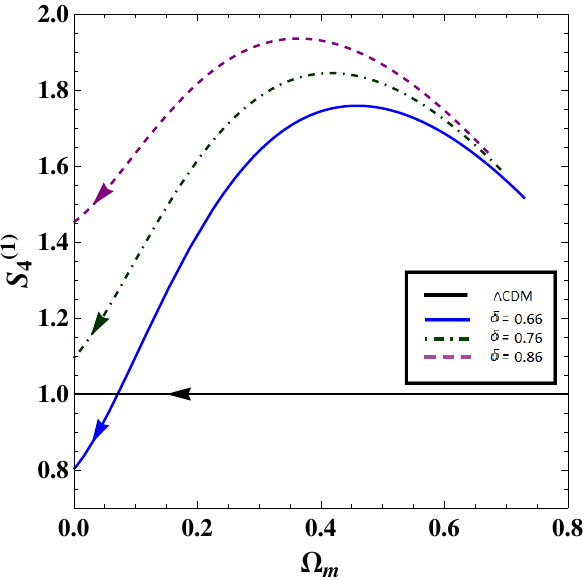}
     \end{minipage}
     \hspace{1cm}
     \begin{minipage}{0.425\textwidth}
         \centering
        \includegraphics[width=\textwidth]{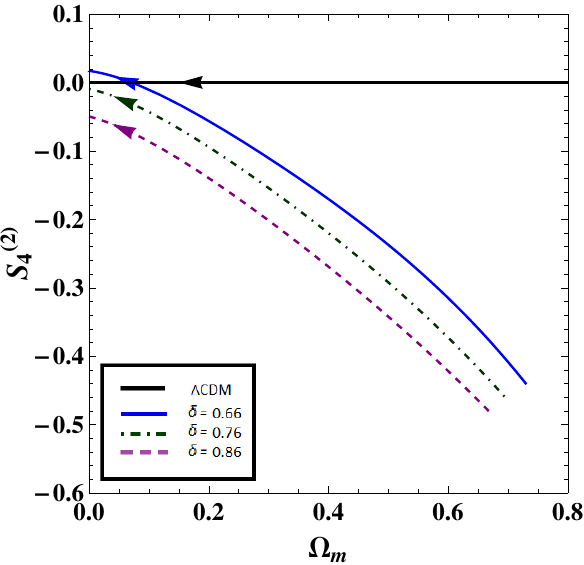}
     \end{minipage}
        \caption{The left and the right panels correspond to the evolutionary paths of $S_4^{(1)}$ and $S_4^{(2)}$, respectively, as a function of $\Omega_m$, for different values of $\delta$. The arrows indicate the direction of cosmic evolution. The initial condition is considered as $\Omega_d(x=-\mbox{ln}(1+z)=0) \equiv \Omega_{d0} \approx 0.7$.}
        \label{fig 3}
\end{figure*}

The evolutionary paths of $S_4^{(1)}$ and $S_4^{(2)}$ as a function of $\Omega_m$ are presented in FIG. \ref{fig 3}. The degeneracy of $S_4^{(1)}-\Omega_m$ curves (left panel) is mostly broken in current epoch ($\Omega_m\approx0.3$) and form  convex vertices in $0 \leq \Omega_m \leq 0.8$ region. All $S_4^{(2)}-\Omega_m$ trajectories (right panel) grow monotonically to $\Lambda CDM$ line in $0 \leq \Omega_m \leq 0.8$ region but they exhibit almost no variation for different values of $\delta$.\\

\begin{figure*}[ht]
     \centering
     \begin{minipage}{0.425\textwidth}
         \centering
         \includegraphics[width=0.98\textwidth]{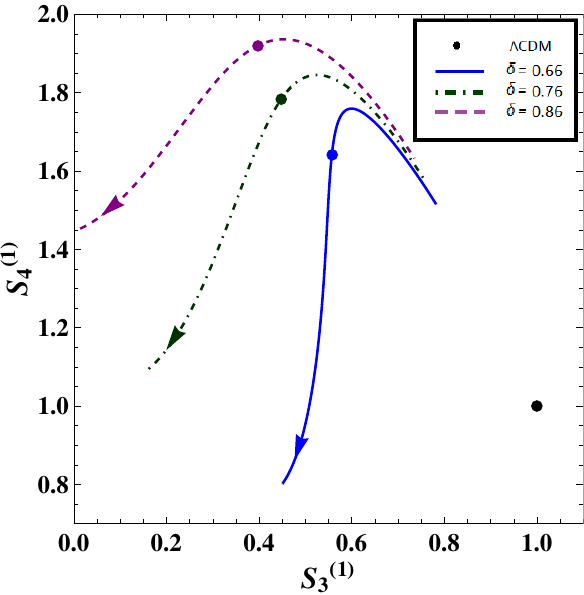}
     \end{minipage}
     \hspace{1cm}
     \begin{minipage}{0.425\textwidth}
         \centering
       \includegraphics[width=\textwidth]{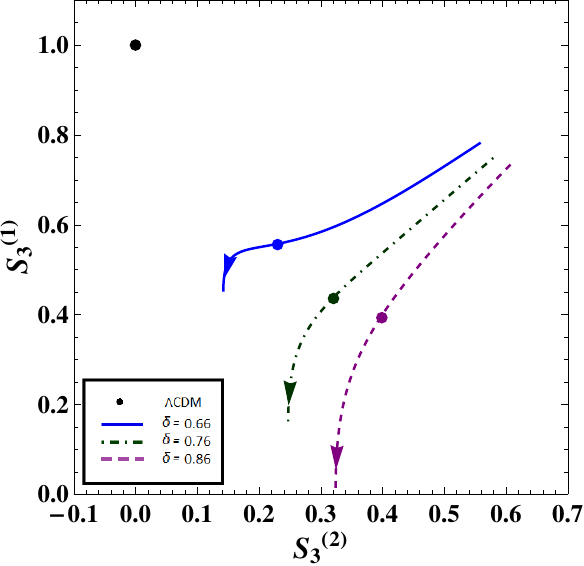}
     \end{minipage}
        \caption{The left and the right panels correspond to the statefinder pairs \{$S_3^{(1)},S_4^{(1)}$\} and \{$S_3^{(2)},S_3^{(1)}$\}, respectively, for  different values of $\delta$. The arrows indicate the direction of cosmic evolution and the dots represent the present values of the parameters. The initial condition is considered as $\Omega_d(x=-\mbox{ln}(1+z)=0) \equiv \Omega_{d0} \approx 0.7$.}
        \label{fig 4}
\end{figure*}

In FIG. \ref{fig 4} we use the Statefinder \{$S_3^{(1)},S_4^{(1)}$\}(left panel) and \{$S_3^{(2)},S_3^{(1)}$\}(right panel) to discriminate curves of non interacting VHDE model for different values of $\delta$. In $S_4^{(1)}-S_3^{(1)}$
plane all the curves evolve near $\Lambda CDM$ point. We observe that for smaller $\delta$, the present value (denoted by dots on the curves) is closer to $\Lambda CDM$ point. In both $S_4^{(1)}-S_3^{(1)}$ and $S_3^{(1)}-S_3^{(2)}$ plane the differences between the present values on evolving curves for various $\delta$ values and the fixed point of $\Lambda CDM$ are fairly evident.

\section{Growth rate of perturbations}\label{sec04}
The fractional growth parameter is defined as \cite{Acquaviva:2008qp,Acquaviva:2010vr}
\begin{equation}
    \epsilon(z) = \frac{f(z)}{f_{\Lambda CDM}(z)}
\end{equation}
In this context, $f(z) = \frac{d\ln \delta_\rho}{d \ln a}$ represents the growth rate of structure. Here, $\delta_\rho = \frac{\delta \rho_m}{\rho_m}$, where $ \delta \rho_m$ and $\rho_m$ denote the density perturbation and the energy density of matter (which includes both CDM and baryons), respectively. Assuming that the perturbation is linear and that there is no interaction between DM and DE, the perturbation equation at late times can be represented as
\begin{equation}
    \Ddot{\delta_\rho}+2H\dot{\delta_\rho}=4\pi\rho_m\delta_\rho
\end{equation}
in gravitational units $G=1$. As a result, the approximate growth rate of linear density perturbations can be approximated as \cite{Wang:1998gt}
\begin{equation}
    f(z)\simeq \Omega_m(z)^\gamma,
\end{equation}
where
\begin{equation}
\gamma(z) = \frac{3}{5 - \frac{w_d}{1-w_d}} + \frac{\left\{3\left(1-w_d\right)\left(1-\frac{3w_d}{2}\right)\right\}\left(1-\Omega_m(z)\right)}{125\left(1-\frac{6w_d}{5}\right)^3}.
\end{equation}
When the EoS parameter $w_d$ is either constant or changes gradually over time, the approximation remains quite accurate. It has been demostrated in Ref. \cite{Arabsalmani:2011fz} that the fractional growth parameter $\epsilon(z)$, when utilized in conjunction with the statefinder hierarchy, can serve as an effective diagnostic tool, represented as the composite null diagnostic, CND $\equiv\{S_n,\epsilon\}$. The growth index in the $\Lambda$CDM model is approximately $\gamma \approx 0.55$, and the value of $\epsilon(z)$ is consistently equal to 1 \cite{Wang:1998gt,Linder:2005in}. As a result, the diagnostic pair $\{S_n,\epsilon\} \equiv \{1, 1\}$. This acts as a reference point or a ``null test'' --- any deviation from this standard indicates behavior that diverges from the $\Lambda$CDM model.

\begin{figure}[!htbp]
\centering
\includegraphics[width=0.425\textwidth]{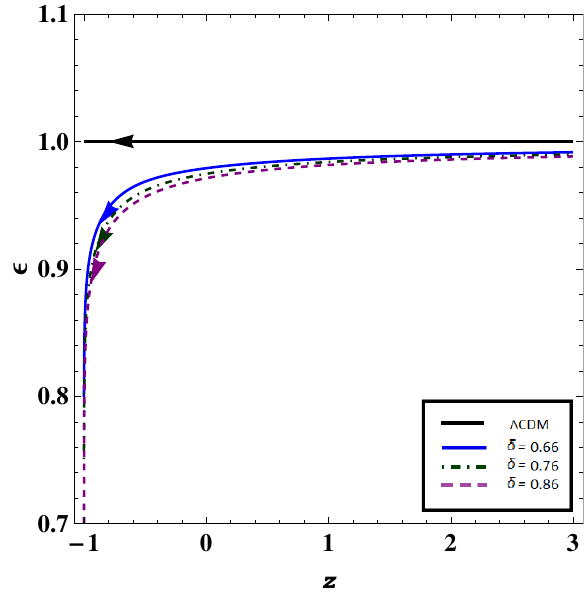}
\caption{The evolutionary paths of $\epsilon$ against the redshift $z$ for different values of $\delta$. The arrows indicate the direction of cosmic evolution. The initial condition is considered as $\Omega_d(x=-\mbox{ln}(1+z)=0) \equiv \Omega_{d0} \approx 0.7$.}
\label{fig 5}
\end{figure}

In FIG. \ref{fig 5}, we illustrate the evolutionary paths of $\epsilon(z)$ as a function of redshift $z$ in the VHDE model, considering different values of $\delta$. It is observed that all the trajectories of $\epsilon(z)$ evolve beneath the $\Lambda$CDM line $\epsilon(z)=1$ and have a trend of coincidence at high redshift zones. 
The curves decrease in the $-1 \leq z \leq 0$ region rapidly and again have a trend of coincidence at low redshift zones. When compared with the illustrations in FIGs. \ref{fig 1}, \ref{fig 2}, and \ref{fig 3}, it is observed that the evolutionary trajectories of $\epsilon$ show significant difference than those of $S_3^{(1)}$ and $S_4^{(1)}$. Therefore, relying solely on the single geometric diagnostic is inadequate. It would be more effective to integrate it with the fractional growth parameter, as a composite null diagnostic (CND), to achieve clearer differentiation \cite{Li:2014mua}.\\

\begin{figure*}[ht]
     \centering
     \begin{minipage}{0.425\textwidth}
         \centering
         \includegraphics[width=\textwidth]{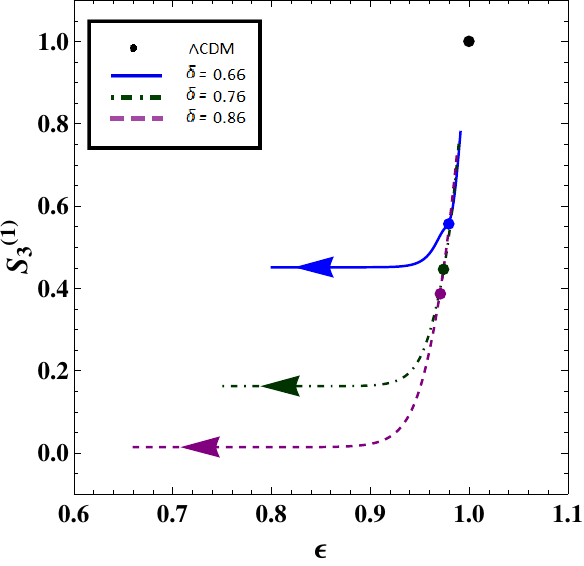}
     \end{minipage}
     \hspace{1cm}
     \begin{minipage}{0.425\textwidth}
         \centering
         \includegraphics[width=\textwidth]{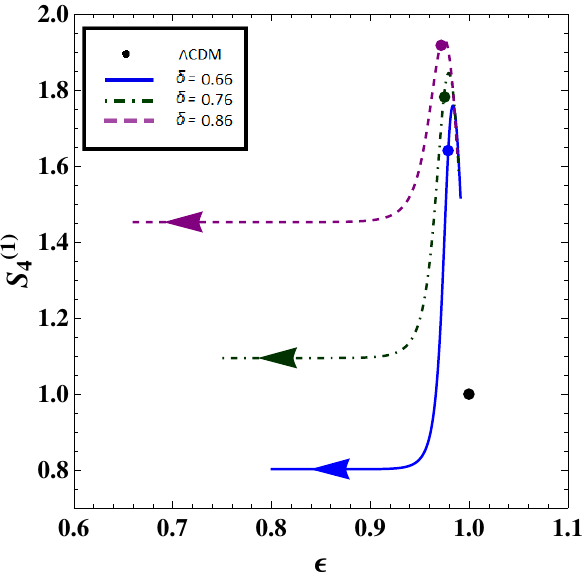}
     \end{minipage}
        \caption{The left and the right panels correspond to the evolution of the CND pairs $\{S_3^{(1)},\epsilon\}$ and $\{S_4^{(1)},\epsilon\}$, respectively, for different values of $\delta$. The current values of $\{S_3^{(1)},\epsilon\}$ and $\{S_4^{(1)},\epsilon\}$ are marked by the colored dots, while the black dot represents those for the $\Lambda$CDM model. The arrows indicate the direction of cosmic evolution.}
        \label{fig 6}
\end{figure*}

The evolutionary paths of $\{S_3^{(1)},\epsilon\}$ and $\{S_4^{(1)},\epsilon\}$ for non interacting VHDE model are illustrated in FIG. \ref{fig 6}, for different values of $\delta$. The present values of $\{S_3^{(1)},\epsilon\}$ and $\{S_4^{(1)},\epsilon\}$ are indicated by round dots, while the $\lambda$CDM fixed point $(1,1)$ is represented by black dots for comparison. The deviations between the VHDE model with various $\delta$ values from the $\Lambda$CDM model is measured by the distances of the corresponding dots from the black dot, represented by the $\Lambda$CDM. It is observed that by applying the CND, $\{S_3^{(1)},\epsilon\}$ and $\{S_4^{(1)},\epsilon\}$, the differences between the evolving trajectories and the fixed point of $\Lambda$CDM are quite apparent.

\section{$\omega_d-\omega_d'$ analysis and stability analysis of the model}\label{sec05}
\begin{figure}[hbt!]
\centering
\includegraphics[width=0.425\textwidth]{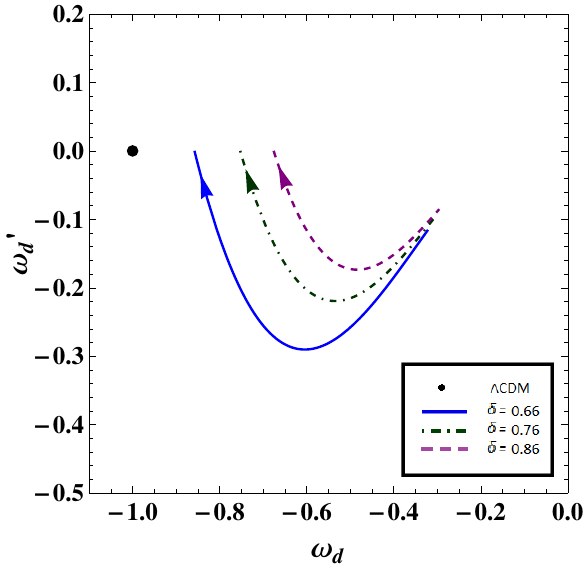}
\caption{The evolutionary paths in $\omega_d-\omega_d'$ plane for different values of $\delta$. The $\Lambda$CDM model, corresponding to $\{\omega_d-\omega_d'\}  = \{-1, 0\}$, is represented by a black dot for comparison.}
\label{fig 7}
\end{figure}

Now we focus on the development of the DE EoS parameter $\omega_d$ and its derivative with respect to $x=\mbox{ln}~a$, expressed as  $\omega_d' = \frac{d\omega_d}{d\ln a}$. We also investigate how different model parameters impact the evolutionary trajectories within the $\omega_d-\omega_d'$ plane. The $\omega_d-\omega_d'$ plane can be categorized into two distinct groups \cite{Caldwell:2005tm,Dubey:2025hsf}. The freezing model is characterized by $\omega_d'<0$ when $\omega_d<0$, indicating an accelerated expansion of the Universe, whereas the thawing model is represented by $\omega_d'>0$ at $\omega_d<0$. In this dynamical analysis, the fixed point $\omega_d=-1$, $\omega_d'=0$ signifies the standard $\Lambda$CDM model on the $\omega_d-\omega_d'$ diagram. FIG. \ref{fig 7} depicts the evolutionary paths in the $\omega_d-\omega_d'$ plane. Notably, these trajectories evolve in the freezing region with $\omega_d'<0$ at $\omega_d<0$ and approaches the standard $\Lambda$CDM model more closely for smaller values of $\delta$.\\

Nest, we analyze the squared sound speed, $v_s^2$, a crucial parameter in the investigation of cosmological models, as it dictates the propagation of perturbations within the cosmic fluid. 
The squared sound speed $v_s^2$ thus governs how pressure responds to density perturbations, directly shaping structure growth and the model's stability \cite{Luciano:2025fox}. Its sign is the key diagnostic: $v_s^2>0$ gives a real sound speed, so perturbations propagate as ordinary waves and remain bounded, a signature of classical stability. If $v_s^2<0$, the sound speed becomes imaginary, and perturbations grow exponentially instead, indicating instability \cite{Kim:2004is,Myung:2007pn,Luciano:2025fox}. Mathematically, the squared sound speed, $v_s^2$, is given as
\begin{equation}
    v_s^2=\frac{\dot{p_d}}{\dot{\rho_d}}=\omega_d-\frac{\omega_d'}{3(1+\omega_d)}.    
\end{equation}

\begin{figure}[hbt!]
\centering
\includegraphics[width=0.425\textwidth]{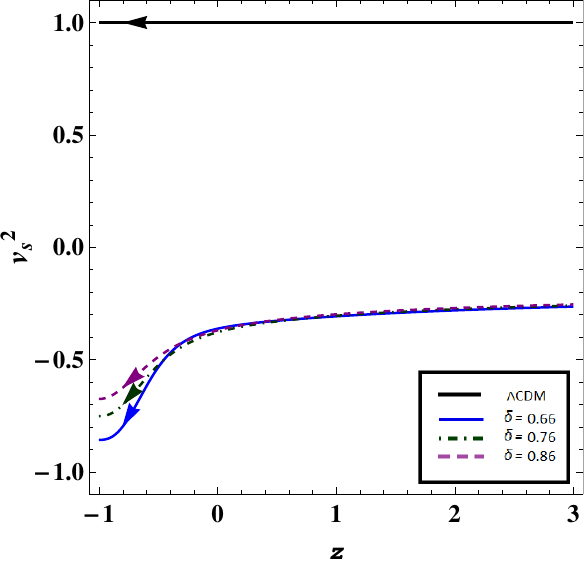}
\caption{The evolutionary paths of $v_s^2$ as a function of $z$, for different values of $\delta$. Arrows indicate the
time directions of evolution. }
\label{fig 8}
\end{figure}

FIG. \ref{fig 8} illustrates the evolution of $v_s^2$ for various values of $\delta$, indicating that $v_s^2$ remains below zero across the entire redshift range examined. In standard cosmological perturbation theory, such a condition often indicates the existence of classical instabilities in scalar perturbations, which tend to grow exponentially and may ultimately make a model unphysical \cite{Dubey:2025hsf}. 
Following Ref. \cite{Luciano:2025fox}, a negative $v_s^2$ over some redshift interval does not necessarily indicate a serious problem with the model. Several points support this fact. Firstly, if the background evolution remains stable and the resulting perturbations stay small, this behavior can be regarded as acceptable within reasonable theoretical limits \cite{Kim:2004is,Myung:2007pn}. Secondly, $v_s^2$ alone does not give the full picture, a complete assessment requires a perturbative analysis that includes both scalar and metric perturbations, since effects such as a modified effective sound speed, entropy perturbations, or non-adiabatic contributions can stabilize the evolution at the perturbative level even when the adiabatic $v_s^2$ is negative. Finally, this behavior is not unique to the present model; Tsallis HDE, for example, shows a similar sign change in $v_s^2$ over a comparable redshift range \cite{Sharif:2019seo}, suggesting that such classical instabilities may be common among extended-entropy HDE reconstructions and merit further investigation into their underlying dynamics and possible stabilizing mechanisms.

\section{CONCLUSIONS}\label{sec07}
This paper focuses on the study of the recently proposed Viaggiu holographic dark energy (VHDE) model \cite{Saha:2026lnb} with regards to statefinder hierarchy and the growth rate of perturbations. 
In addition, the composite null diagnostic (CND) pair $\{S_n,\epsilon\}$, which combines these two elements, is also frequently employed to assess DE models. In VHDE, there exists only one independent parameter $\delta$, which we adjust to explore the impacts on this model, making use of these diagnostic tools, and differentiate it from the $\Lambda$CDM model. Moreover, we assess model stability by analyzing the squared speed of sound $v_{s}^{2}$ and also perform an $\omega_d-\omega_d'$ analysis to further explore the evolutionary behavior of the DE component.
 Employing $S_4^{(1)}$, which includes the fourth order derivatives of the scale factor, provides a more pronounced differentiation between the VHDE models for different values of $\delta$ and the $\Lambda$CDM model, as opposed to the outcomes obtained from $S_3^{(1)}$. Our analysis in the $S_3^{(1)}-\Omega_m$, $S_3^{(2)}-\Omega_m$, $S_4^{(1)}-\Omega_m$ and $S_4^{(2)}-\Omega_m$ planes demonstrates that the VHDE model is non-degenerate around the current $\Omega_m$ value ($\approx 0.3$), for different values of $\delta$. We find that with smaller $\delta$, the present value (indicated by dots on the curves) aligns more closely with the $\Lambda$CDM point. In the $S_4^{(1)}-S_3^{(1)}$ and $S_3^{(1)}-S_3^{(2)}$ planes, the variations between the present values on the evolving curves for different $\delta$ values and the fixed point of $\Lambda$CDM are clearly observable. The evolutionary paths in the $\{S_3^{(1)}, \epsilon\}$ plane display unique characteristics and the deviations from $\Lambda$CDM can be accurately assessed. Consequently, the statefinder hierarchy, which includes higher derivatives of the scale factor, along with the CND, is quite useful in distinguishing between VHDE models and the $\Lambda$CDM model, and the CND is highly effective in resolving degeneracies for different values of the $\delta-$parameter in the VHDE model. The geometric diagnostic $\omega_d-\omega_d'$ reveal that the curves do not intersect the $\omega_d=-1$ (phantom divide) line and the trajectories evolve from the freezing region $(\omega_d < 0, \omega_d' < 0)$ towards the conventional $\Lambda$CDM behavior in future. Furthermore, our analysis reveals that the squared sound speed $v_s^2$ is negative, indicating instability of the current model under perturbations, which is a considerable limitation in its existing formulation. We are optimistic that future, high-precision observations, like those from SNAP-type studies, will be able to precisely determine cosmological parameters, thereby enhancing our understanding of the properties associated with the VHDE model.

 \subsection*{Declaration of Generative AI and AI-assisted technologies}
 During the preparation of this work, the author used Claude (Anthropic) and ChatGPT (OpenAI) to assist with phrasing and clarity in portions of the text that discuss and contextualize relevant prior literature. After using these tools, the author reviewed and edited the content as needed and takes full responsibility for the content of the published work.
 
\begin{acknowledgements}
The author is grateful to Sree Chaitanya College, Habra and Jadavpur University for extending all necessary administrative support during his Ph.D. research work. The author expresses sincere gratitude to Professor Subenoy Chakraborty, Dr. Nilanjana Mahata, Dr. Subhajit Saha, and Dr. Abdulla Al Mamon for their insightful comments on the ﬁrst draft of this manuscript.
\end{acknowledgements}

\end{document}